\documentclass{IEEEtran}

\usepackage[pdftex,bookmarksnumbered,colorlinks,bookmarks,breaklinks,linktocpage, citecolor=black, linkcolor=black]{hyperref}
\usepackage{graphicx}
\usepackage{algorithm}
\usepackage{amsmath}
\usepackage{graphicx}
\usepackage{hyperref}
\usepackage[noend]{algpseudocode}
\usepackage{enumitem}
\newcommand{\eat}[1]{}

\begin{document}


\title{Improving virtual host efficiency through resource and interference aware scheduling}


\author{
\IEEEauthorblockN{Evangelos Angelou,  Konstantinos Kaffes, Athanasia Asiki, Georgios Goumas and Nectarios Koziris} \\
\IEEEauthorblockA{Computing Systems Laboratory, School of Electrical and Computer Engineering\\
National Technical University of Athens\\
\{eangelou, kkaffes, aassiki, goumas, nkoziris\}@cslab.ece.ntua.gr
}
}

\maketitle

\begin{abstract}
Modern Infrastructure-as-a-Service Clouds operate in a competitive environment that caters to any user's requirements for computing resources. The sharing of the various types of resources by  diverse applications poses a series of challenges in order to optimize resource utilization while avoiding performance degradation caused by application interference.
In this paper, we  present two scheduling methodologies enforcing consolidation techniques  on multicore physical machines. Our resource-aware and interference-aware scheduling schemes aim at improving physical  host efficiency while preserving the application performance by taking into account host oversubscription and the resulting workload interference. We validate our fully operational framework through a set of real-life workloads representing  a wide class of modern cloud applications. The experimental results prove the efficiency of our system in optimizing resource utilization and thus energy consumption even in the presence of oversubscription. Both methodologies achieve significant reductions of the CPU time consumed, reaching up to 50$\%$, while at the same time maintaining workload performance compared to widely used scheduling schemes under a variety of representative cloud platform scenarios.
\end{abstract}


\vspace{-.1in}

\section{Introduction}\label{sec:intro}
Infrastructure-as-a-Service (IaaS) Clouds have been adopted as a dominant
computing model in IT infrastructures offering ubiquitous access to huge
pools of computing resources.  Numerous user communities, spanning from
early adopters and elasticity-seeking service providers, to the World's
biggest commercial and government agencies, harvest the benefits of elastic,
on-demand resource provisioning in a cost-efficient manner. Thus,
a major challenge in large-scale infrastructures is how to efficiently manage different types
of available resources to increase
utilization without sacrificing performance.

Resource provisioning by IaaS providers is performed with the use of
virtualization technologies that enable the sharing of a physical server among
several workloads. The placement of multiple virtualized workloads in the
same physical host, hence \emph{workload consolidation}, can significantly
contribute to higher availability of physical resources under numerous user
requests and energy savings \cite{Sub2013, Gan2009} (consequently leading
to lower operational costs for datacenter (DC) operators), a very critical
aspect considering that power consumption of such data centers is growing
at an unprecedented rate. Workload consolidation can potentially achieve
better resource utilization in IaaS clouds mainly due to the diversity
and over-estimation of application resource demands. Cloud applications are
coming from a wide variety of domains and present diverse execution profiles,
i.e., they pose different resource demands on the various system components
(CPU, memory, I/O, etc.). Moreover, cloud customers tend to
overestimate the requirements of their applications for computing resources in
an attempt to ensure the expected Quality of Service (QoS) even in worst-case
scenarios; a practice stemming from their experience of in-house infrastructure
provisioning but unnecessary in modern IaaS due to elasticity. Several works
\cite{Beloglazov2012b, DBLP:conf/vee/WangISCQM15, Novakovic2013b} have proposed
resource allocation techniques based on workload consolidation to optimize
resource utilization, increase DC capacity and reduce energy consumption.



Despite the significant benefits, workload consolidation increases
the risk of performance anomalies compared to the execution in
an \emph{isolated} environment, especially for latency critical
workloads \cite{Leverich2014a}. Performance degradation mainly
occurs due to \emph{interference} of  Virtual Machines (VMs) in the same physical host
\cite{Delimitrou2013b}. Although isolated in its own virtual hardware sandbox,
each VM is competing with its neighbors for access to hardware components
that cannot be exclusively provisioned by the host operating system. The
co-allocation of applications can lead to a variety of scenarios ranging
from smooth co-existence\eat{, where application performance is similar to
that of isolated execution,} to catastrophic symbiosis, where one or more
applications suffer from significant performance degradations. To make things
even worse, resource contention may arise unexpectedly at any time between
co-located VMs, severely affecting the offered QoS.  
Several works \cite{Delimitrou2013b, Nath2013, Delimitrou}
propose scheduling schemes attempting to optimize workload co-existence,
while avoiding application interference based on profiling of the considered
applications and their interactions.

In this work we further elaborate upon contention-aware scheduling by
additionally considering \emph{oversubscription}, either as a mechanism to
further increase the utilization of system resources or as a challenging
execution scenario \cite{Bas2012, Hous2014} that needs to be tackled by the
system's resource allocator. We introduce two scheduling approaches \emph{on
a per host level} that utilize oversubscription and enforce consolidation
taking into account the system architecture, the application behavior
and the diversity of workloads. Our goal is twofold: (i) to increase the
core utilization of resources and reduce energy consumption and (ii) to
minimize the impact of interference on performance for latency critical
workloads. Towards these directions, we propose a lightweight solution that
can be fully integrated in a cloud platform to provide efficient scheduling by
estimating \emph{online} the expected workload interference and consolidating
the `\emph{best-fit}' candidates together. The grouping of such workloads
is performed during their execution in our scheduling approaches, since we
introduce: (i) a Resource-Aware Scheduler (RAS) that decides which workloads
to consolidate together based on real-time resource utilization metrics,
without requiring any prior knowledge of the application's behavior and (ii)
an Interference-Aware Scheduler (IAS) exploiting knowledge extracted from
a minimum co-execution profiling setup of some generic workload pairs in
order to determine groups of workloads with minimum interference.

The introduced scheduling strategies exploit dynamic pinning for
multicore systems, a useful practice for heavily-loaded cluster systems
\cite{Podzimek2015}, to coallocate workloads with less interference and minimize the number of underutilized CPUs. Both schemes do not
rely unilaterally on operating system level metrics (e.g., CPU utilization,
memory capacity demand, I/O utilization demand) but take also into account
microarchitecture-level resource interference considering memory bandwidth.
\eat{The scheduling process in both cases evolves both in \emph{time}, since
scheduling decisions are refined over time and in \emph{space}, as dynamic
pinning enables the direct assignment of workloads to specific cores and an
increased degree of isolation among different cores.} Finally, we implement
and evaluate the proposed scheduling techniques on state-of-the-art hardware
using diverse real-life workloads, demonstrating increased efficiency of
15-30\% at a performance cost of no more than 10\% compared with simpler but
widely used schedulers, and present simultaneous performance and efficiency
improvements of up to 10\% and 30\% respectively in oversubscribed hosts.

The paper is organized as follows: Section \ref{sec:related}
outlines previously published work related to our approach; Section
\ref{sec:architecture} presents the architectural details of our
implementation; Section \ref{sec:scheduling}  discusses in detail our scheduling approaches, with the relevant experimental results presented and discussed
in Section \ref{sec:experiments}; we conclude in Section \ref{sec:conclusions}
with lessons learned and further extensions of our approach.

\section{Related Work}\label{sec:related}
Given the importance of VM workload consolidation in modern IaaS, it is no surprise that a large number of works deals with this problem.
Quasar \cite{Delimitrou} increases resource utilization in IaaS clusters by combining user-imposed constraints on application performance and classification techniques to predict application impact on computing resources. This information is used to efficiently pack workloads on available resources, while continuous performance monitoring  may trigger quick resource allocation adjustments. Although our scheduler also monitors and dynamically reallocates workloads, it operates in the context of each compute host. It also aims at improving resource utilization in each host, even in oversubscribed scenarios, while reducing energy consumption.

Paragon \cite{Delimitrou2013b} considers the impact of co-scheduled application interference when scheduling VMs in large, heterogeneous datacenters.  Although our scheduling algorithms are also based on avoiding application interference, they also take into account CPU cores as a shared VM resource. Novakovic et al. also describe DeepDive\cite{Novakovic2013b}, a system that manages VM interference in IaaS clouds, and evaluate its performance. However, DeepDive does not deal with resource oversubscription and its monolithic scheduler design presents latency problems in large cluster management.

Roytman et al. tackle the problem of VM consolidation in PACMan \cite{Nath2013}, minimizing the average degradation of workloads' performance, mainly due to interference, vs energy efficiency. Evaluation of their system with SPEC CPU 2006 benchmarks shows that PACMan realizes 30\% savings in energy costs and up to 52\% reduction in performance degradation compared to consolidation approaches that do not consider degradation. However, given their experimental setup, performance degradation is only measured in terms of CPU utilization, while real-world workloads may suffer degradation due to the sharing of other host resources. Furthermore, the PACMan scheduler is not evaluated in oversubscribed environments, as in our case.

Leverich and Kozyrakis \cite{Leverich2014a} quantify the impact to latency-sensitive workload performance for co-located applications in a shared cluster environment. They show that, contrary to the traditional view, some latency-critical workloads can be co-located to improve DC resource utilization, and still achieve good QoS, using a variety of techniques. In our analysis, we also treat carefully latency-critical applications' performance during co-scheduling. Taking into account the main causes of performance degradation as presented in \cite{Leverich2014a}, namely queuing delay, scheduling delay, and thread load imbalance, our implementation addresses queuing and scheduling delay for all workloads by pinning workloads to specific resources, whilst thread load imbalance is orthogonal to our approach.

The treatment of oversubscribed environments is at an early stage with most efforts \cite{7238586}, \cite{7285284} attempting to identify and quantify candidate bottlenecks of shared resources (e.g., network resources) for performance degradation. In \cite{Baset2012a}, Baset et al. discuss the practice of oversubscription in clouds and detail some of the theory behind it. They discuss VM overload detection and VM migration and VM shutdown as methods to alleviate it, but they do not take advantage of interference detection to efficiently schedule VMs in oversubscribed hosts.

Kannan et al. \cite{Guo2007a} present a number of prototypes to alleviate co-scheduled VM demands on shared resources of on-chip multiprocessor (CMP) platforms. By following their suggestions on proper management of shared last-level cache among co-scheduled VMs, scheduling algorithms can be designed that do not suffer from non-deterministic performance degradation. Although such degradation may happen when workloads are dependent on memory bandwidth, our approach currently focuses on general-purpose workload co-scheduling.


Apart from generic schedulers in cloud infrastructures, various efforts such as \cite{Delgado2015}, \cite{Schwarzkopf2013} focus on specific types of workloads, e.g., map-reduce jobs.  For instance,  Omega\cite{Schwarzkopf2013} is a shared-state, optimistic, transaction-based scheduler that appears as an attractive platform for development of specialized schedulers. To illustrate its flexibility, a MapReduce scheduler is added with opportunistic resource adjustment that benefits 50$-$70\% of MapReduce jobs. Although these approaches are intriguing, they do not address VM scheduling but take into account only a particular workload type (variable CPU-time map-reduce jobs). They demonstrate little improvement in the over-subscribed case relative to centralized schedulers.

Finally, Podzimek et al. \cite{Podzimek2015} experiment with various CPU-pinning strategies of different VM and LXC-container workloads. They conclude that less common CPU pinning configurations (such as ``per-chip'' for heavily loaded systems) improve energy efficiency at partial background loads, indicating that systems hosting co-located workloads could benefit from dynamic CPU pinning based on CPU load and workload type. In our work, we observe similar performance and energy efficiency variation, and we adopt an architecture-neutral approach for our CPU-pinning strategy, which we do not claim as optimal, but sufficient and general enough to be applied to a large variety of commodity hardware based DCs.
 
\section{Architecture}\label{sec:architecture}

\begin{figure}[t]
 \centering\includegraphics[width=1.8in]{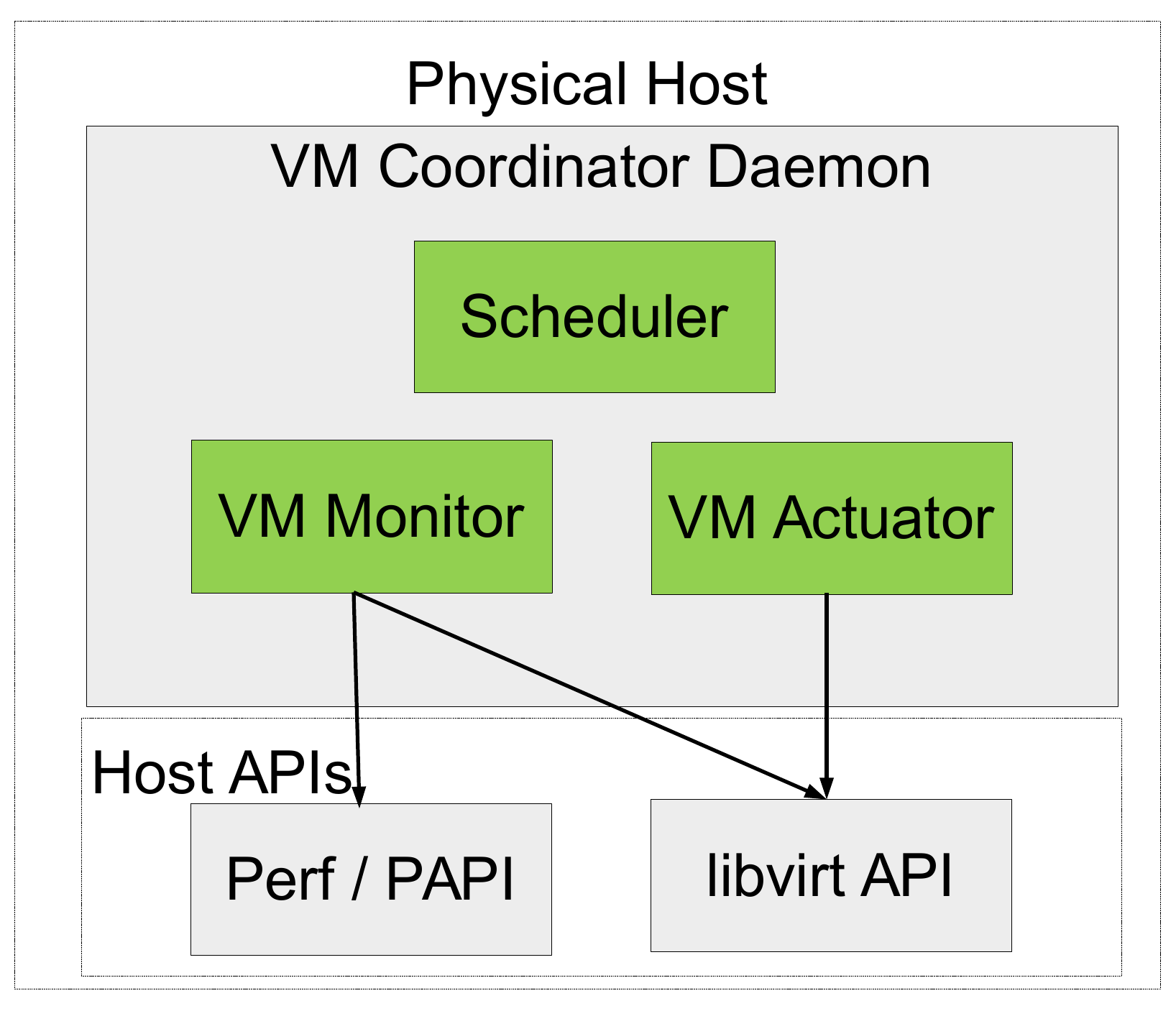}
	\caption{VM Coordinator Daemon architecture.}\label{daemon_arch_internal}
 \vspace{-.2in}
\end{figure}



As previously discussed, most attempts to consolidate workloads focus on exploiting VM migration to alleviate the strain on oversubscribed servers. However, these approaches fail when the infrastructure as a whole is oversubscribed and thus VM migration is technically unreliable and proportionately more expensive in terms of migration time and resource usage.

In our work, we propose a less complex approach to reduce the effects of VM interference on over-subscribed servers. Our intuition is that instead of relying on a global reshuffle of VM workloads across all DC servers, a local optimization approach for each host would reduce workload interference and contention for resources, with less overhead and consequent pitfalls of a centralized scheduler which expects complete infrastructure knowledge. We also hypothesize that by improving resource usage and application performance in each server, the whole cluster's operation is optimized, without the need for complex monitoring and scheduling.

To validate the proposed techniques, we developed a userspace system daemon in Python that schedules guest VMs on available host resources (e.g., CPU cores) depending on a variety of algorithms. Our \emph{VM Coordinator daemon (VMCd)} is composed of three basic modules as shown in Figure \ref{daemon_arch_internal}, namely: (i) a \emph{VM Monitor} module that periodically polls the hypervisor through libvirt for VM resource usage and also uses low-level access to system performance counters through the \texttt{perf} utility, (ii) a \emph{VM Actuator} that can manage VMs and their placement on host cores through the libvirt API and (iii) a \emph{Scheduler} component that can be used to abstract particular placement policies for VMs. In detail, the VMCd modules are:


\begin{description}[leftmargin=0pt,labelindent=0pt]
\item[VM Monitor:]
Monitoring is a critical process for predicting the behavior of workloads so as to optimally consolidate them. The parameters used and monitored per VM in our analysis are: (i) CPU utilization, (ii) DiskIO, (iii) NetIO and (iv) Memory Bandwidth utilization as percentages of the total system resources. The VM Monitor uses the libvirt API\cite{libvirt} to gather CPU pinning and then CPU, DiskIO and NetIO utilization of the running VMs. To calculate the consumed Memory Bandwidth of a socket and the per VM Memory Bandwidth Utilization as proposed in \cite{DBLP:conf/vee/WangISCQM15}, we use the \texttt{perf} events presented in Table \ref{tab:perf_counters}. 

\item[VM Actuator:] The VM Actuator provides a high-level abstraction to libvirt API calls to perform the necessary actions required by the Scheduler. It can manage VMs throughout their life-cycle and enforce the required CPU pinning adjustments.

\item[Scheduler:] The Scheduler module is an implementation of the algorithms described in Section \ref{subsec:scheduling}. As VMCd's core component, it operates by optimizing the pinning of VMs on physical hosts at regular intervals. To accomplish this, it obtains a list of the idle and running workloads on the server. We consider a workload to be idle if its CPU usage during the last monitoring time window was below $2.5\%$. Using the VM Actuator, every idle workload is pinned on a specific server core and considered to consume zero resources. Then, the running workloads are pinned on the rest of the server's cores according to the pinning selection algorithm implemented. The scheduler's architecture is depicted in Algorithm \ref{alg:GeneralScheduler}.
\end{description}

\begin{algorithm}
\floatname{algorithm}{}
\caption{General Scheduler}\label{alg:GeneralScheduler}
\begin{algorithmic}[1]
\Procedure{Schedule:}{}
\While {\texttt{True}}
  \State \texttt{sleep timeInterval}
  \State \texttt{idleWloads} $\gets$ \texttt{GetIdleWloads()}
  \State \texttt{runningWloads} $\gets$ \texttt{GetRunningWloads()}
  \For {\texttt{Wload in idleWloads}}
    \State \texttt{PlaceVM(0,Wload)}
  \EndFor
  \For {\texttt{Wload in runningWloads}}
    \State \texttt{targetC} $\gets$ \texttt{SelectPinning(Wload)}
    \State \texttt{PlaceVM(targetC,Wload)}
  \EndFor
\EndWhile
\EndProcedure
\end{algorithmic}
\end{algorithm}

\begin{table}
\centering
\caption{Performance Counters} \label{tab:perf_counters}
\begin{tabular}{|c|c|}
 \hline
 \textbf{Hardware Events} & \textbf{Description}\\
 \hline
UNC\_QMC\_NORMAL\_READS & Memory Reads\\
 \hline
 UNC\_QMC\_NORMAL\_WRITES & Memory Writes\\
 \hline
 OFFCORE\_RESPONSE & Requests serviced by DRAM\\
 \hline
\end{tabular}
 \vspace{-.2in}
\end{table}

VMCd resides on each physical compute host and dynamically monitors VM resource usage, as previously described. Upon detecting a possible action for VM consolidation, VMCd modifies the placement of VMs on the available host cores by altering their CPU pinning through libvirt. Also, as new workloads are forwarded to VMCd, they are pinned to CPU cores as resource availability allows. In both cases, VMCd takes into account information provided by the user about the type of workload (e.g. CPU-intensive, latency-critical) to decide how to co-schedule VMs. Although we don't target NUMA architectures or distributed memory architectures specifically, our design can easily accommodate them.


\section{Scheduling Policies}\label{sec:scheduling}
\subsection{Workload Profiling}\label{subsec:profiling}

In order to determine the resource utilization and pairwise interference of each workload, an offline profiling phase takes place for each type of workload, before workload scheduling takes place. During profiling, each workload is run isolated and co-pinned to the same core with every other workload, in order to derive the interference in each possible pairwise combination. Taking into account the slowdown of each workload when pinned on the same core with every other one, we produce $S$, a $N\times N$ matrix, assuming $N$ distinct workloads (hence also referred as jobs) representing $N$ application classes:

\begin{equation}
S_{ij}(\Psi) = \frac{P(\psi_i,\psi_j)}{P(\psi_i)}
\end{equation}
where $\Psi$ the set of the $N$ distinct classes $\psi_1,\psi_2,...,\psi_{N}$ and $P$ the measured performance of the workload. $P$ corresponds to the most significant performance metric for the application type, e.g. completion time for CPU-intensive workloads or latency in ms for latency-critical applications. This provides a realistic measure of the performance that a user experiences and would probably use to define QoS from her point of view. 

Next, we obtain a set of metrics consisting of the CPU, DiskIO, NetIO and Memory Bandwidth utilization of every workload as a percentage of the total server's resources when using it in isolated mode. If there are $N$ jobs or workload classes, the percentage of resource utilization of each workload is represented by a $N \times M$ matrix $U$, where $M$ is the number of monitored metrics (four in our case).

The initial profiling phase is used to inform the scheduler, when slowdown $S$ and resource utilization $U$ matrices are determined. These profiles are considered static throughout the execution of each workload, as most VM workloads' resource demands are determined by the nature of the user application. Thus, during scheduler operation, we assume that each workload is tagged according to its profile. Although this tagging is external to the scheduler, we expect it to be performed by the user in an attempt to ensure QoS for her particular application.

\subsection{Scheduling algorithms}\label{subsec:scheduling}

Scheduling is a fundamental process and an intrinsic part of our daemon. The Scheduler's job is to place incoming workloads on the cores of a server using limited information about them. We assume that the set of jobs that must run on the server is predefined and cannot change regardless of the overhead they introduce, i.e., interserver workload scheduling is not considered as it is orthogonal to our approach. Once the datacenter management system assigns a set of VMs to a server, our scheduler takes over and pins them to the available cores according to a policy. We have designed, implemented and evaluated two scheduling algorithms: The first one aims to maximize resource utilization so as to increase throughput per core; the second one takes into account the expected interference of the incoming workloads and tries to find a suitable balance between performance and resource efficiency.

Both approaches try to accomplish two discrete but complementary goals. If the server is undersubscribed, they try to consolidate workloads with minimal performance degradation. Thus, the first goal is to save cores so as to assign new jobs on them or allow the cores to revert to their lowest power state. The second goal refers to the case of oversubscription, where the algorithms try to determine the optimal placement that minimizes performance degradation induced by workload co-location. The proposed approaches, namely \emph{Resource-Aware Scheduler} and \emph{Interference-Aware Scheduler}, are presented in the following subsections.

\subsubsection{Resource-Aware Scheduler (RAS)}
The first scheduler takes decisions based on resource utilization by the host's workloads using the resource utilization matrix $U$. Let us assume that there is a core $c$ with a set of $k$ active workloads, $A_c = \{a_1,..,a_k\}$ whose resource utilization is described by matrix $U_c$ consisting of the subset of matrix $U$ with the utilization values of only the active core workloads $A_c$. $U$, and therefore $U_c$, are derived during the profiling phase defined in Subsection \ref{subsec:profiling} to define the core overload $OL_c(A)$:

\begin{equation}
OL_c(A_c) = \sum\limits_{\substack{j=1}}^M \max(0,\sum\limits_{\substack{i=1}}^k U_c[i,j] - thr)
\end{equation}

This metric considers only the composite load beyond the resource utilization threshold $thr$ for every resource consumed by the active VMs pinned on the core. Our intuition is that once a particular resource has been allocated up to a certain threshold, it is then inefficient to further pin VMs on that particular core, irrespective of the type of resource that they mostly use (e.g., CPU, DiskIO, NetIO or Memory Bandwidth). Thus, this approach will try to match workloads with different execution profiles (i.e., requiring different resources) and proceed this way until the defined threshold is reached.

\begin{algorithm}
\floatname{algorithm}{}
\caption{Resource Aware Scheduler}\label{alg:ras}
\begin{algorithmic}[1]

\Procedure{SelectPinning(Wload):}{}
\For {\texttt{i in range(cores)}}
  \If {$OL_i(A_i \cup Wload) = 0$} 
    \State \Return \texttt{i}
  \EndIf
\EndFor

\State \texttt{minUsage} $\gets$ $OL_0(A_i \cup Wload) - OL_0(A_i)$
\State \texttt{minCoreID}   $\gets$ \texttt{0}

\For {\texttt{i in range(1,cores)}}
  \State \texttt{temp} $\gets$ $OL_i(A_i \cup Wload) - OL_i(A_i)$
  \If {\texttt{temp<minUsage}}
    \State \texttt{minUsage} $\gets$ \texttt{temp}
    \State \texttt{minCoreID}   $\gets$ \texttt{i}
  \EndIf
\EndFor\\
~~~~\Return \texttt{minCoreID}
\EndProcedure 
\end{algorithmic}
\end{algorithm}

Towards this, the scheduler uses a matrix of the pinning of all running jobs that have already been placed on the server and the total utilization of each resource on each core. When the decision for the placement of a new VM workload $a_{k+1}$ is taken, the scheduler checks if there is a core, $c$, whose overload  after the new job is added, $OL_c(A_c \cup a_{k+1})$, is zero. If such a core exists, the workload's virtual CPU is pinned on that core. If not, the scheduler, having determined the load of every core on the server with and without the new workload, then places it to the core whose overload $OL_c$ will increase the least with the new job. These steps are outlined in Algorithm \ref{alg:ras}.

After the placement, the resource usage is updated for the core on which the workload is pinned  according to its profile in matrix $U$, i.e., the CPU utilization of the selected core, the Memory Bandwidth usage for all cores in the same socket and the NetIO and DiskIO usage for all cores in the server. This information is kept to determine the degree of resource oversubscription for each server resource, as in this case the monitoring information can only be used to determine the fraction of resources actually allocated to each VM, but not the fraction of resources requested by each workload. Therefore, this map contains the amount of resources that would be ideally acquired by the co-pinned VMs and used to determine the sum of resource contention exerted among workloads.

In our experiments we have set the value of \emph{thr} equal to $120\%$. We have derived this value during the initial classification, since this value is sufficient to allow workload co-location without significant degradation of workload performance. In a sense, this parameter determines the aggressiveness of the scheduler with regard to VM consolidation and we plan to experiment further with different values of this parameter in the future. An exhaustive examination however is out of scope for our comparison in this particular case.

A simpler version of RAS is also formulated taking into account only one metric, the CPU utilization of incoming workloads. We name it \emph{CPU Aware Scheduler (CAS)} and we use it as a reference point in our experiments.

\subsubsection{Interference-Aware Scheduler (IAS)}

Another approach to optimize resource utilization is to consider the resultant interference among co-located workloads on the same core. In our analysis, we first explore space-sharing placements by pinning the workloads on the cores of a physical host to share concurrently all host resources \emph{with the exception of CPU cores}. Since it is common for workloads not to operate at peak load, we extend our approach by adopting time-sharing of the CPU itself. We therefore consider co-located VMs that are concurrently pinned on the same CPU core making CPU a time-shared resource. Nevertheless, in this case the VM workloads would suffer also from CPU interference, which is similar to other resource interference, and stems from multiple core context-switches as multiple VM workloads contend for CPU time on the same core. The goal of the IAS is to address the above challenges and minimize co-located workload interference due to space and time sharing, when VM consolidation is performed.

To implement IAS, we assume knowledge of the slowdown each job suffers when pinned on the same CPU with every other job, i.e., represented by the measures of matrix $S$ as calculated during the profiling phase (see Section \ref{subsec:profiling}). For each workload $a_i$ in the list of $k$ active workloads $A_c$ placed on a CPU core $c$, and with $S_c$ the matrix containing pairwise slowdown for each job class placed on core $c$, we define as workload interference for $a_i$ on core $c$, $WI_{a_i}(A_c)$:

\begin{equation}
WI_{a_i}(A_c) = \frac{\sum\limits_{\substack{j=1 \\ j\neq i}}^k S_c[i,j] + \prod\limits_{\substack{j=1 \\ j\neq i}}^k S_c[i,j]}{2}
\end{equation}



where matrix $S_c$ is derived from matrix $S$ and contains the pairwise slowdown ratio of all VM workloads on a particular core, expressed as an absolute number. 

This criterion can estimate the interference caused by more than two co-located workloads in an accurate manner, as we consider infeasible the on-line or off-line measurement of the interference among multiple workloads (in case that more than two workloads are pinned on the same core). On-line determination of the exact interference would involve running all variations of workloads pinned on any subset of the cores for a fixed time, a calculation that not only scales exponentially with the number of workloads and host CPU cores, but would also incur a severe performance penalty on all running workloads. Off-line exhaustive profiling would suffer from runtime variations due to workload behavior or host heterogeneity. Our estimation of multiple workload interference from off-line pairwise workload profiling is easy to implement and provides enough accuracy for our purposes. Thus, compared to other approaches where each possible workload set is evaluated on-line\cite{Nath2013} and resources are used for an exhaustive search before scheduling, we do not require any prior knowledge of performance degradation of each VM when consolidated with any set of other VMs. Having just the one-by-one slowdowns available, we estimate the slowdown for a specific workload when consolidated with multiple other VMs.

Our formula also overestimates the expected average slowdown between workloads by incorporating the product of all the co-located workloads' slowdown. From the calculation of $WI_{a_i}(A_c)$ above we derive that the slowdown product acts differently depending on the slowdown value: it provides a smaller fraction of expected interference when the slowdown is sub-linear (i.e., $<2.0$) and exponentially increases after that point. This estimation tries to penalize the co-location of heavily interfering workloads over the co-location of well behaving workloads. Thus, IAS will try to keep apart heavily interfering workloads, even if it has to  co-locate a larger number of lightly interfering workloads on some particular cores. We expect that this behavior will attain an acceptable level of performance degradation for both types of workloads.

When calculating the expected interference of a workload, our formula takes the average of the sum and the product of the slowdowns the workload will suffer when co-located with each of the workloads already placed in the server. We use this complex value instead of a simple formula (e.g. just the product or the sum) as it helps avoid irrational predictions in some cases. For example, let's assume that in a server which hosts three workloads, a new job arrives which has expected slowdown S equal to 1 with all of them. On the one hand, if we used just the product of the slowdowns as a metric, the calculated interference the workload would suffer if placed in that server, would be 1. This prediction in most cases would be a stark underestimation as there will be four workloads running in the same server and some contention will definitely occur. On the other hand, if the sum of the slowdowns was used, the expected interference would be 3, a very high value for a workload that seems to be insensitive to the already hosted workloads, at least to a certain degree. In that case, our formula would calculate the expected interference to be 2, a value which empirically seems to be closer to the observations than the aforementioned extreme predictions.
\begin{algorithm}
\floatname{algorithm}{}
\caption{Interference-Aware Scheduler}\label{alg:ias}
\begin{algorithmic}[1]

\Procedure{SelectPinning(Wload):}{}
\For {\texttt{i in range(cores)}}
  \If {$I_i(A_i \cup Wload) < threshold$} 
    \State \Return \texttt{i}
  \EndIf
\EndFor

\State \texttt{minInter} $\gets$ $I_0(A_0 \cup Wload)$
\State \texttt{minCoreID}   $\gets$ \texttt{0}

\For {\texttt{i in range(1,cores)}}
  \State \texttt{temp} $\gets$ $I_i(A_i \cup Wload)$
  \If {\texttt{temp<minInter}}
    \State \texttt{minInter} $\gets$ \texttt{temp}
    \State \texttt{minCoreID}   $\gets$ \texttt{i}
  \EndIf
\EndFor\\
~~~~~\Return \texttt{minCoreID}
\EndProcedure 
\end{algorithmic}
\end{algorithm}

We also define the core's interference, $I_c$, to be the maximum of the interferences of all workloads on that core:
\begin{equation}
I_c(A_c) = \max\limits_{\substack{i=1}}^n WI_{a_i}(A_c)
\end{equation}
This formula is used to consider the worst-performing workload on the core, or in other words, the one that suffers maximum interference. It is used to avoid situations where a workload is co-scheduled with workloads that interfere with it heavily, a scenario that would lead to starvation for one or more co-located workloads. 

When the decision for the placement of a new workload $a_{k+1}$ is made, the scheduler checks if there exists a core, $c$, whose interference after the new  workload is placed $I_c(A_c \cup a_{k+1})$ is below a given interference threshold. If this is the case, the workload's virtual CPU is pinned on that core. If no such core exists, the scheduler first determines the interference of every core on the server with the new workload and then pins it on the core whose interference is minimum
after the placement. The placement decision steps are detailed in Algorithm \ref{alg:ias}. We have selected $1.5$ as the $threshold$ used in this case, as close to the average slowdown of a pair of random co-scheduled workloads determined in matrix $S$ during the profiling phase. More formally:
\begin{equation}
threshold  \approx \frac {\sum\limits_{\substack{i=1}}^N  \sum\limits_{\substack{j=1}}^N S[i,j]} {N^2}  
\end{equation}
 
\section{Experiments}\label{sec:experiments}
\begin{figure*}[t]
\centering
\begin{tabular}{cc}
\hspace{-.2in}
 \begin{minipage}{17pc}
        \centering\includegraphics[width=17pc]{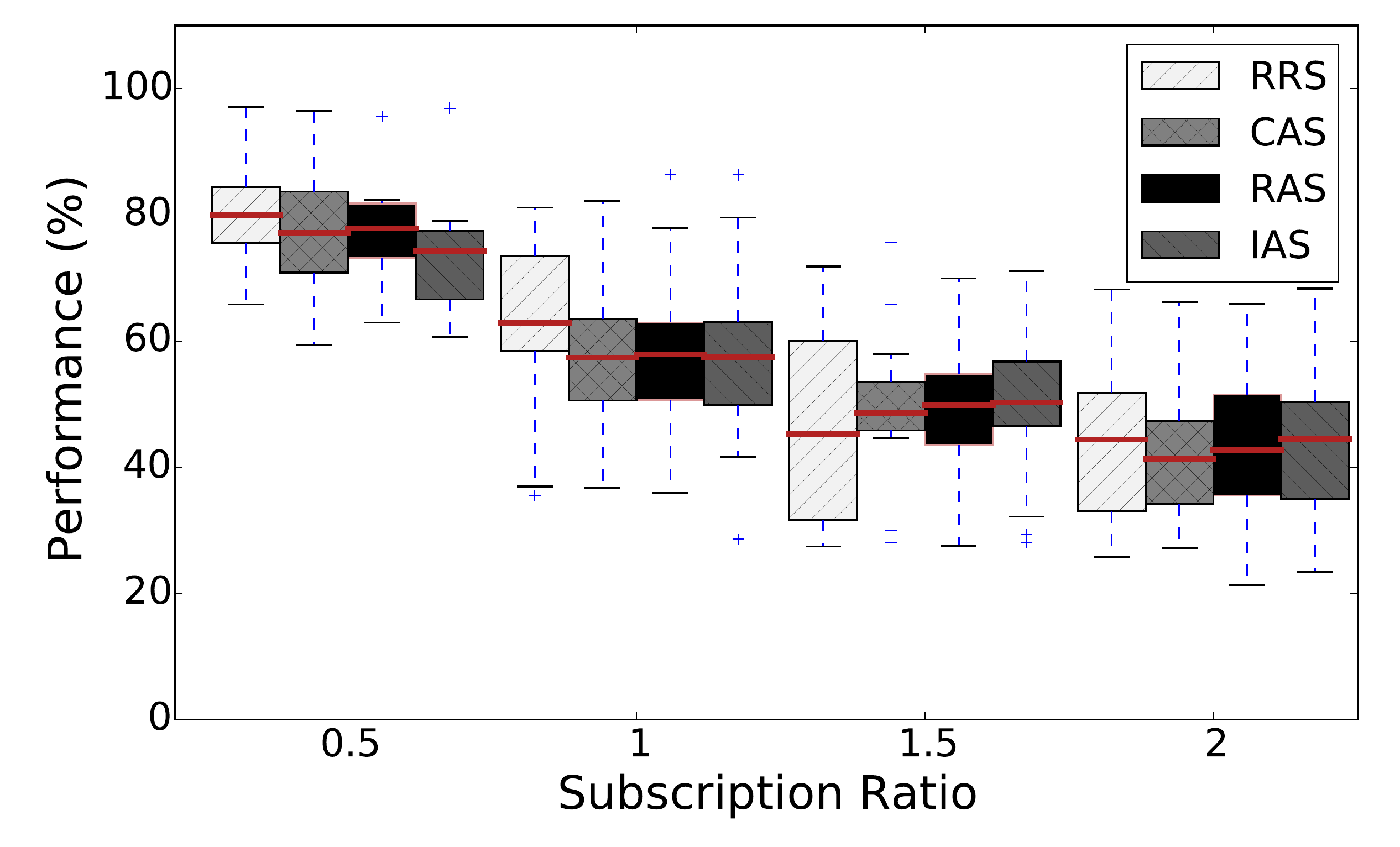}
 \end{minipage}
&
   \begin{minipage}{20pc}
        \centering\includegraphics[width=20pc]{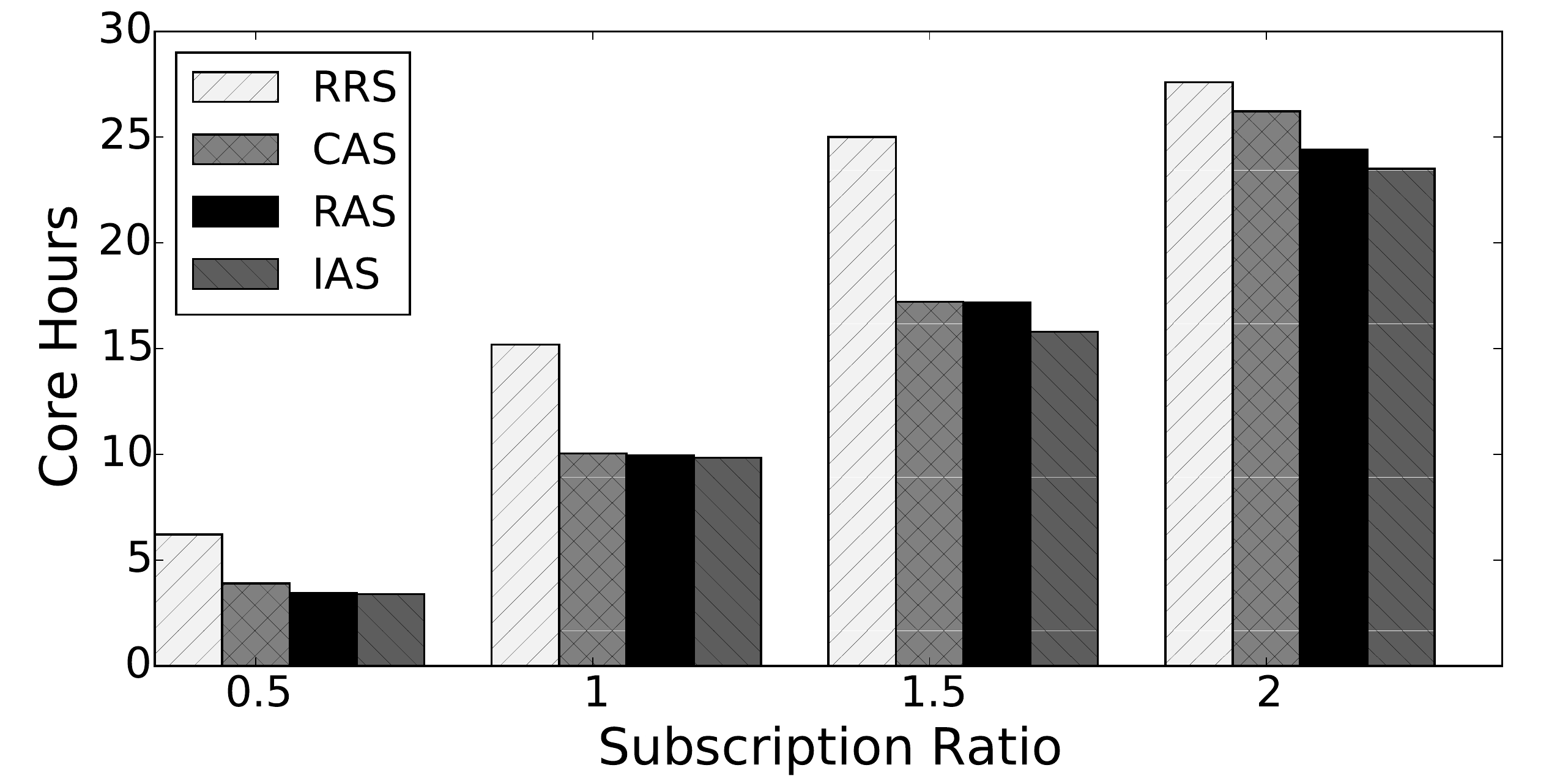}
 \end{minipage}
\end{tabular}
 \vspace{-0.1in}
 \centering\caption{\label{fig:boxperf_RS} Workloads' performance and CPU time consumed for each scheduler: RRS, CAS, RAS and IAS in the Random Scenario.}
\vspace{-.1in}
\end{figure*}

\subsection{Experimental Setup}
Our experimental testbed consists of a single server with two Intel Xeon X5650 Processors. The server has twelve 2.66-GHz cores, divided into two six-core sockets that share 12 MB of LLC. The server also features 48 GB of DRAM and one 1-Gb network port. The following paragraphs summarize the benchmarks and the three scenarios used to evaluate the different schedulers. Scenarios include batch and latency-critical workloads which are large consumers of resources on private and public clouds. A different VM is created for each benchmark. Furthermore, we assume that all VMs have a single virtual core which is pinned to a real core for simplicity.

\begin{figure*}[t]
\centering
\begin{tabular}{cc}
  \hspace{-.2in}
 \begin{minipage}{17pc}
        \centering\includegraphics[width=17pc]{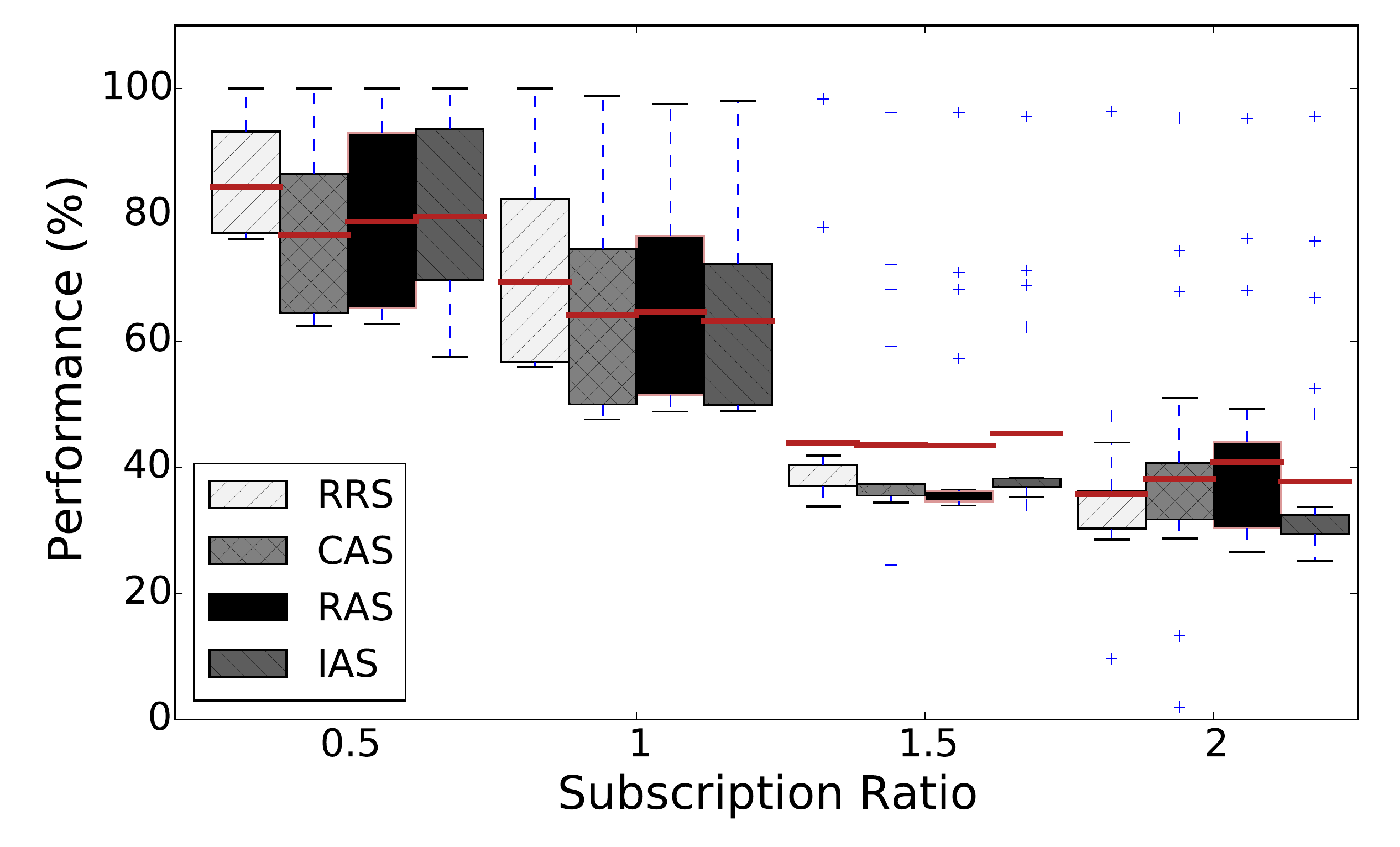}
  \end{minipage}
&
 \begin{minipage}{20pc}
        \centering\includegraphics[width=20pc]{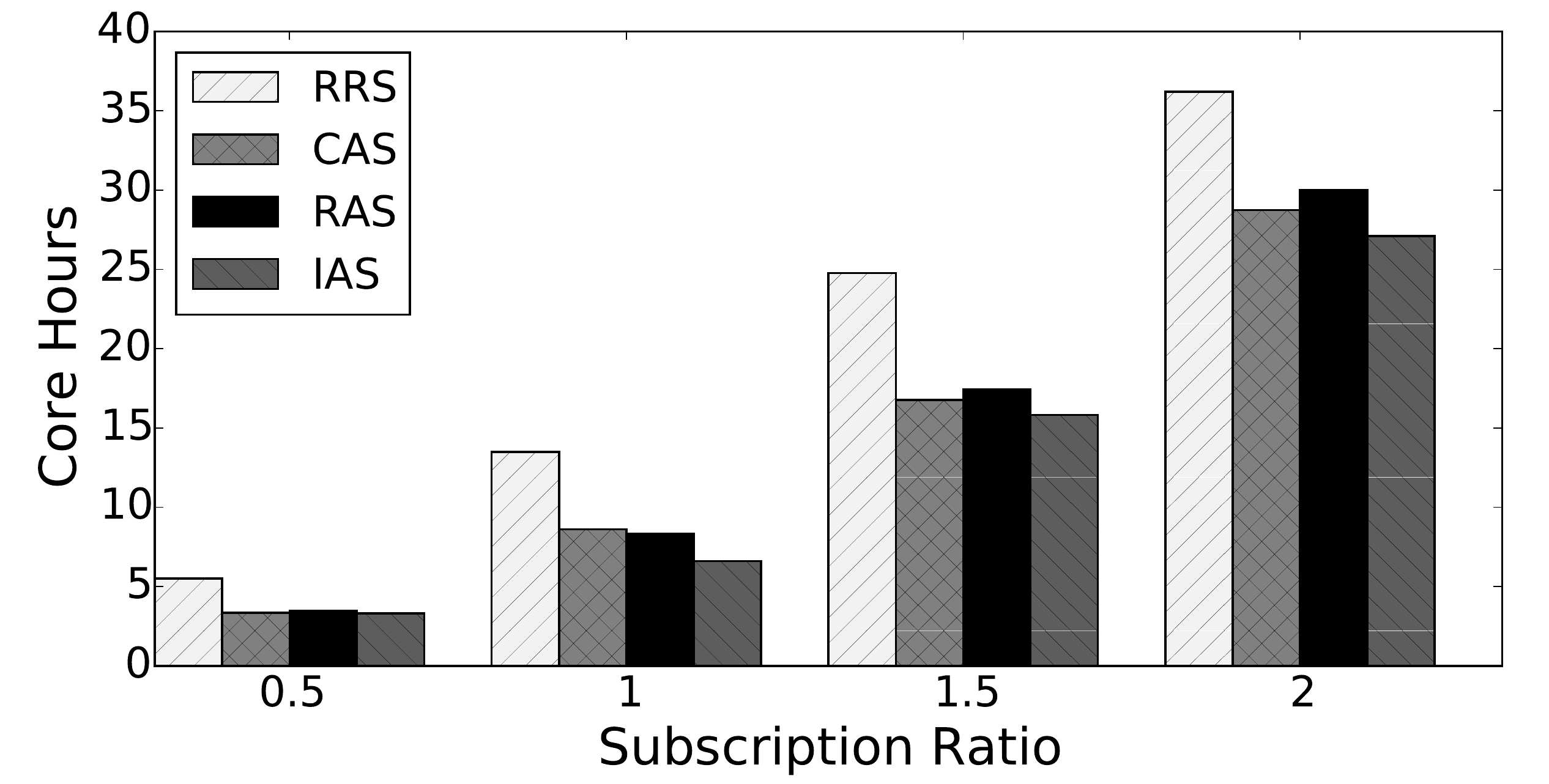}
  \end{minipage}
\end{tabular}
 \vspace{-0.1in}
\centering \caption{\label{fig:boxperf_LC} Workloads' performance and CPU time consumed for each scheduler: RRS, CAS, RAS and IAS in the Latency Critical Scenario.}
\vspace{-.1in}
\end{figure*}

\subsection{Experimental workloads}
To evaluate our implementation, we use a wide variety of common cloud workloads as follows:

\begin{description}[leftmargin=0pt,labelindent=0pt]
\item[Blackscholes:] It represents the wide range of HPC, CPU-intensive workloads and consists of a PDE solver that solves the Black-Scholes equation, designed to eliminate financial risk in the stock derivatives' market. The program's performance is limited by the number of FLOPS the processor can execute. The implementation used in our experiments was procured from the PARSEC suite of benchmarks \cite{bienia09parsec2}.

\item[Hadoop:] The Hadoop analytics framework \cite{hadoop} provides Terasort, a benchmark used to simulate load on a map-reduce cluster. Hadoop terasort is used in our experiments to model analytics workloads running on IaaS Clouds.

\item[Jacobi:] Jacobi runs a 2D stencil computation kernel provided by the PolyBench/C benchmarking suite\cite{polybench} to simulate CPU and memory bandwidth intensive HPC workloads.

\item[LAMP:] To represent web services, we used a simple web server based on Apache to develop a lightweight PHP REST service that can accept GET and PUT requests for random data stored in a MySQL database. We have used Apache JMeter \cite{jmeter} to stress the server; a performance measuring framework. Using JMeter, two access patterns were defined to simulate heavy and light load by varying the number of client threads.

\item[Media Streaming:] To simulate a video streaming server we used the CloudSuite \cite{Ferdman:2012:CCS:2150976.2150982} media streaming module consisting of two components, the Darwin Streaming Server and an RTSP client used to stress it. We have created three different versions of this workload, simulating low, medium and high server load by varying the number of concurrent client threads.
\end{description}

To measure workload performance, we use appropriate metrics depending on the nature of each benchmark. In the case of batch workloads (Blackscholes, Hadoop, Jacobi), we evaluate performance using workload run time, while for LAMP and Media Streaming we measure the number of requests served per second and the server throughput in kbps respectively.

\begin{figure*}[t]
\centering
\begin{tabular}{cc}
\hspace{-.15in}
 \begin{minipage}{20pc}
        \centering\includegraphics[width=19pc]{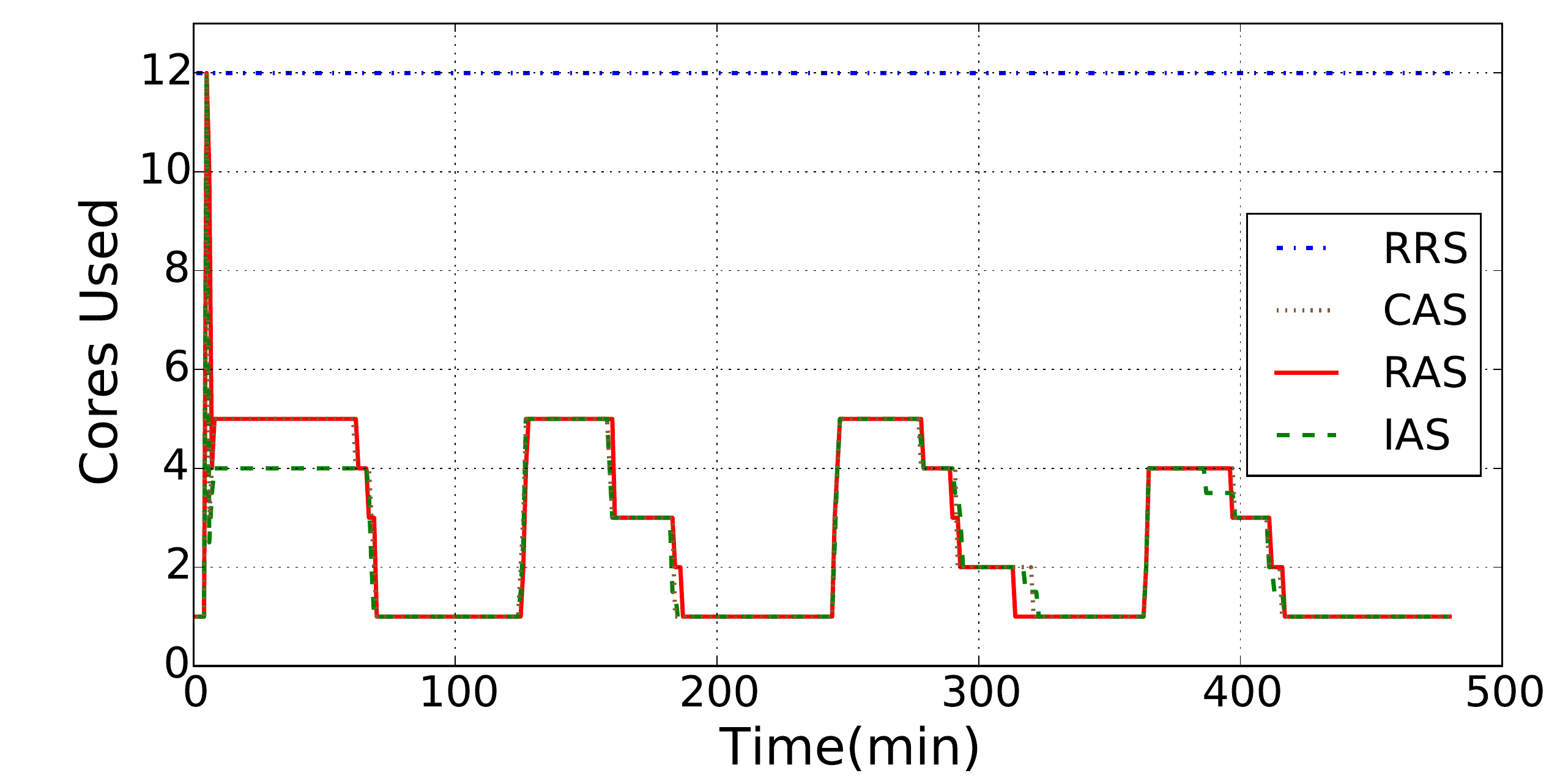}
         \vspace{-.15in}
        \caption{Time series of CPU consumption for the 6-batch scenario.}\label{coresDS6}
 \end{minipage}
&
 \begin{minipage}{20pc}
        \centering\includegraphics[width=19pc]{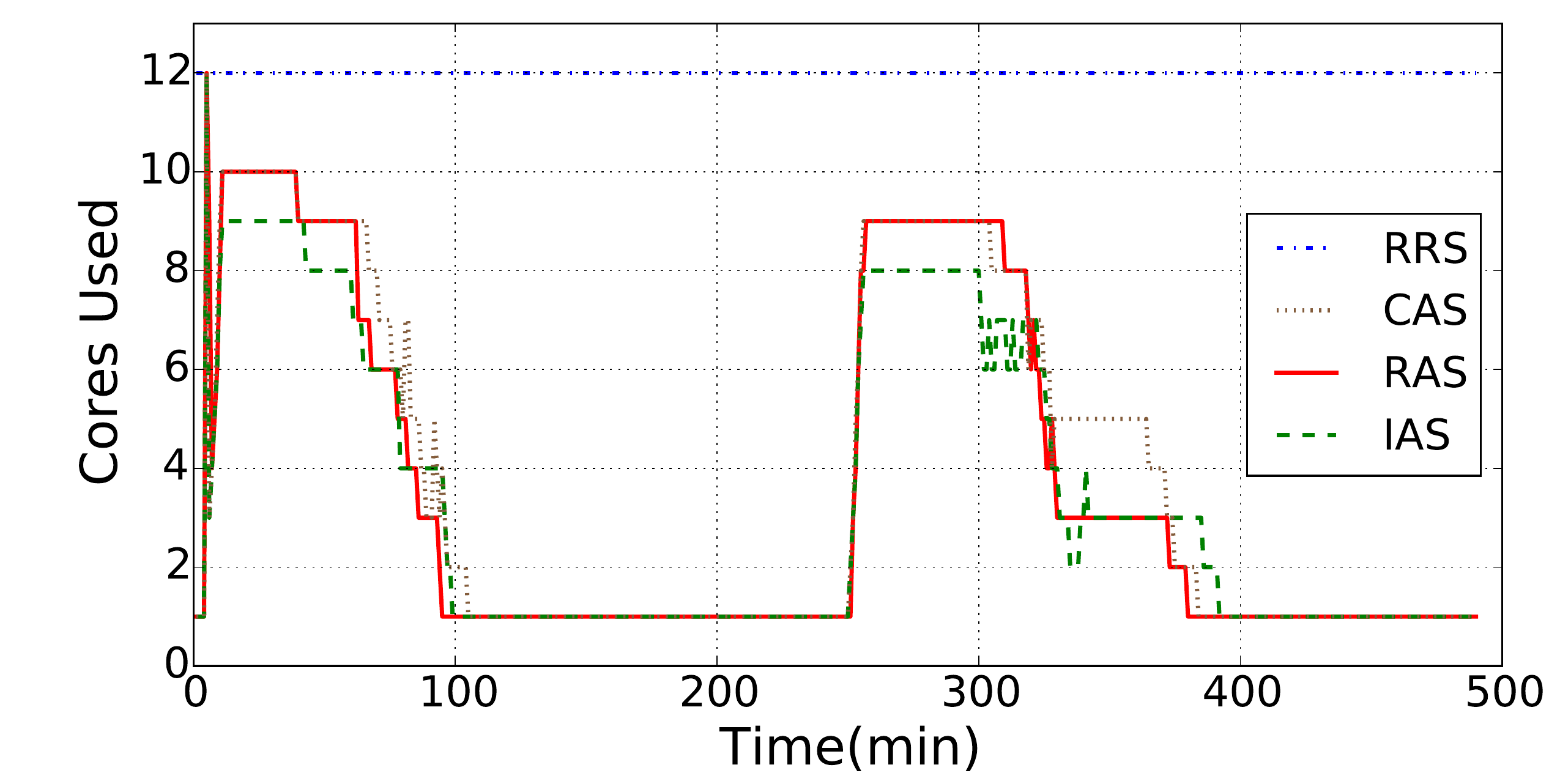}
        \vspace{-.15in}
        \caption{Time series of CPU consumption for the 12-batch scenario.}\label{coresDS12}
  \end{minipage}
 \end{tabular}
\vspace{-0.2in}
\end{figure*}

\subsection{Experimental Scenarios}

\subsubsection{Random Scenario}

The first scenario examined is a random scenario of all workload types. The server is shared between batch, media streaming and latency critical benchmarks. This scenario is used to evaluate the schedulers' performance under unexpected conditions. Workloads arrive with 30 seconds inter-arrival time. We define as \emph {Subscription Ratio ($SR$)} the ratio of the number of jobs placed in the server to the number of cores the server has, i.e., in our 12-core experimental server a subscription ratio of 0.5 signifies the placement of 6 jobs on the server while a ratio of 2 indicates that 24 workloads have been placed.

We measure the average performance of all scenario workloads compared to their performance in an isolated environment and the total CPU hours consumed by all workloads until scenario completion. In our analysis, we compare the achieved performance and resource utilization with the widely used \emph{Round Robin Scheduler (RRS)} that iterates over the list of workloads, pinning each workload in sequence on a different core. RRS is interference and resource unaware, and unable to detect whether a workload is in running state or idle, since it does not take note of the monitoring metrics. \eat{With regards to CPU time consumption by the RRS, we assume that throughout the runtime of a scenario, all cores hosting a job are active even if the VM that hosts the job is idle. The rationale for this is that the CPU cannot revert to the lowest power state as long as there are even small number of instructions originating from the hypervisor thread or the guest OS that have to be executed on the pinned CPU.}

The results of the random scenario are shown in Figure \ref{fig:boxperf_RS}. When the server is undersubscribed, i.e., $SR \leq 1$, CAS, RAS and IAS should result to large savings in core hours suffering only minimal performance degradation over RRS. The experimental results confirm this expectation as for $SR=0.5$ RAS saves on average $44\%$ in cores hours with only $2.5\%$ performance degradation over RRS, while IAS also saves around $45\%$ in core hours suffering though $7\%$ performance degradation. For $SR=1$, IAS performs better, reducing by $34.5\%$ on average the core hours consumed, while workload performance degradation suffered is $8.5\%$, with RAS offering a little worse results. In both cases CAS performance is similar to RAS confirming the fact that in the set of workloads that we use, CPU is the most important performance bottleneck.


For $SR=1.5$, we observe improvements for RAS and IAS over RRS both in core hours ($30\%-35\%$) and performance ($10\%$), CAS provides similar consumption gains but with less performance improvement.  However, for $SR=2$ the performance is about stable with the core hours improving $11.5\%$ (RAS) to $15\%$ (IAS). This behavior can be explained by the fact that in this case the server is severely oversubscribed and so workload placement does not matter as much in terms of performance, while the small improvements to CPU hours consumption can be attributed to the detection and consolidation of idle workloads. CAS, being oblivious to a number of potential sources of interference, has minimal savings in core hours while performing much worse than RRS.


\subsubsection{Latency-critical Heavy Scenario}
Latency-critical services represent a large percentage of user workloads in cloud platforms. In this scenario, we model such a distribution of workloads by considering a large number of latency-critical but low load applications and a small number of batch and media streaming workloads. Latency-critical services are more sensitive both to time- and space- sharing interference compared to batch and media streaming benchmarks. 
Performance results for CAS, RAS and IAS compared to RRS are presented in Figure \ref{fig:boxperf_LC}. Due to the low CPU load, it is possible for CAS, RAS and IAS to consolidate jobs in less cores than RRS for $SR$ up to $1.5$. This leads to a significant reduction in core hours consumption of at least $30\%$ and up to $50\%$ for IAS in $SR=1$, with performance degradation never exceeding $10\%$. For $SR=2$, CAS and RAS attain a $20\%$ reduction in core hours consumed, together with a $6\%$ (CAS) to $14\%$ (RAS) improvement in performance by avoiding the co-scheduling of workloads that suffer from interference. IAS is more aggressive and provides $30\%$ improvement in CPU time, along with $5.5\%$ performance improvement over RRS.


\subsubsection{Dynamic Scenario}

\begin{figure}[t]
\hspace{-.2in}
\centering\includegraphics[width=0.47\textwidth]{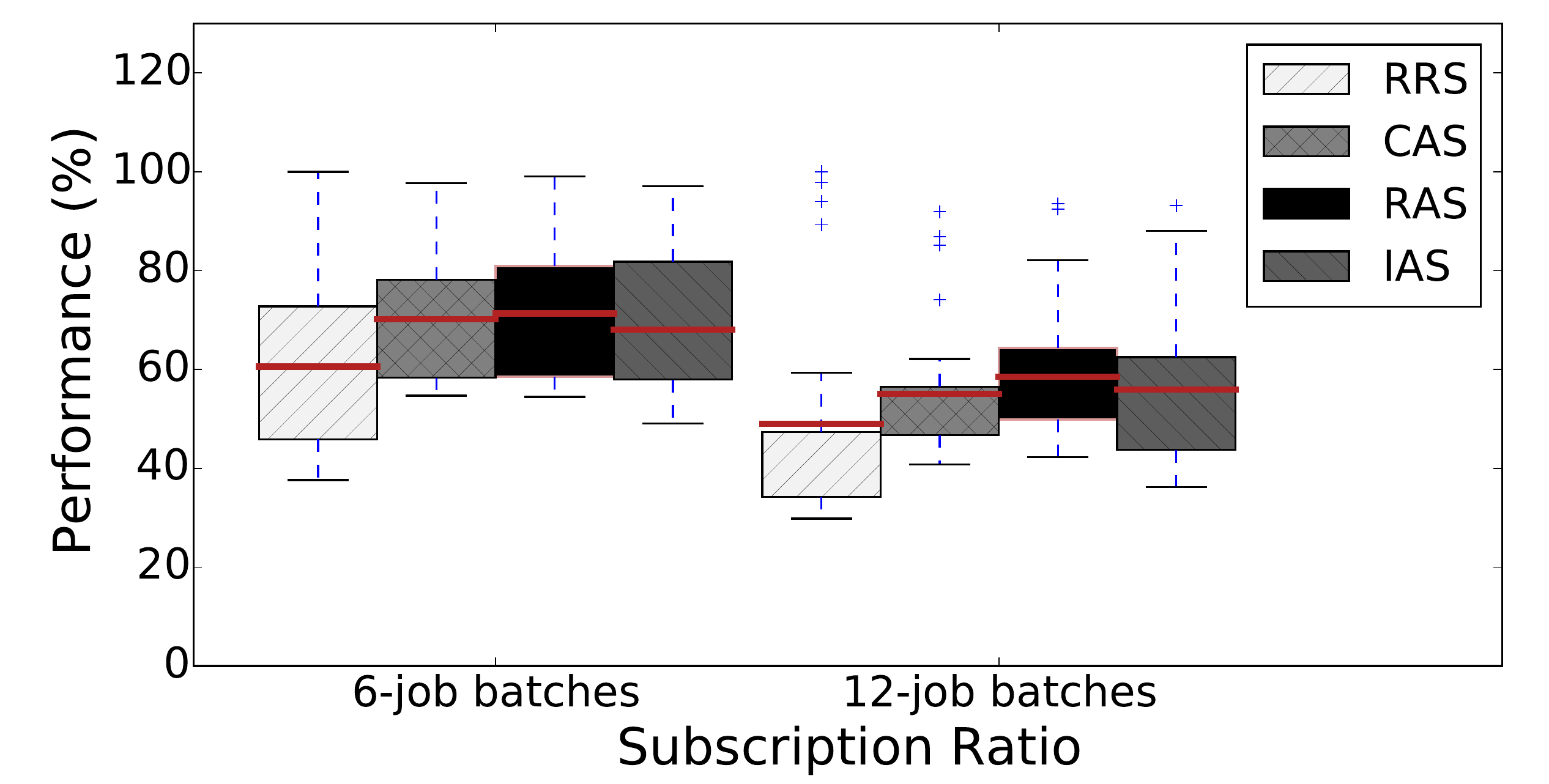}
 \vspace{-0.15in}
\caption{ Performance of workloads for job batches of the dynamic scenario.}\label{fig:boxperfDS}
\vspace{-0.2in}
\end{figure}

The utilization of resources by the same application may vary during its execution according to its execution phase (development or production phase) or load (e.g., in the case of a web service). In order to model time-varying load behavior we designed a scenario where 24 random VMs are placed in the server where they become active in 12- or 6-job batches. The results are displayed in Figures \ref{coresDS6}, \ref{coresDS12} and \ref{fig:boxperfDS}. RRS, being unaware of the monitoring system's metrics and making static decisions about the pinning, needs to reserve the whole server continuously regardless of VMs' state (idle/running). On the other hand, RAS achieves $18\%$ improvement in performance by avoiding the time-sharing of active and sensitive workloads while releasing a large number of cores due to the detection and consolidation of idle workloads. IAS consolidates workloads even more aggressively using less cores than RAS but achieves a smaller performance improvement of $13\%$. CAS exhibits the lower performance of IAS with the the higher core consumption of RAS thus being the least effective of the schedulers.


\section{Conclusions}\label{sec:conclusions}
In this paper, we introduce a resource-aware and an interference-aware scheduling scheme for workload consolidation on the host level of cloud infrastructures.  Our approaches treat the global, DC-level VM consolidation problem as a set of discrete optimizations for the placement of VM workloads on each physical host. The proposed schemes are evaluated compared to the widely used round robin scheduling policy under three realistic scenarios of workload co-existence in a host for different oversubscription rates. The achieved performance is improved in terms of preserving VM QoS and decreasing overall host utilization. We also examine the effects of VM oversubscription on workload QoS and show that by taking into account workload interference both host efficiency and VM performance can be improved.
Moreover, our experimental results show that VM consolidation can be promising in terms of avoiding performance degradation even for latency-critical applications, if appropriately performed.

Further study of resource-aware and interference-aware schedulers for larger subscription ratios is planned in order to validate the savings observed in our experiments to a wider range of scenarios. We also plan to further explore local vs global consolidation approaches using a private cloud to pit our approach against infrastructure-scale schedulers.

\section*{Acknowledgments}
This  work  is  partially supported  by  the  EU project DOLFIN no. 609140 (FP7-SMARTCITIES-2013,
ICT-2013.6.2)\footnote{Dolfin Project, http://www.dolfin-fp7.eu.}.

\bibliographystyle{IEEEtran}
\bstctlcite{IEEEexample:BSTcontrol}
\tiny
\bibliography{IEEEtran,bibliografia}
\normalsize

\end{document}